\documentstyle[prc,preprint,aps,12pt]{revtex}
\begin{document}
\draft
\title {Behavior of the giant-dipole resonance in $^{120}$Sn and $^{208}$Pb\\
at high excitation energy}
\author{W.E.~Ormand$^{a,b}$, P.F.~Bortignon$^{a}$, R.A.~Broglia$^{a,c}$, 
and A.~Bracco$^{a}$}
\address{
$^{a}$Dipartimento di Fisica, Universit\`a Degli Studi di Milano, and \\
Istituto Nazionale di Fisica Nucleare, Sezione di Milano,\\
Via Celoria 16, 20133 Milano, Italy\\
$^{b}$Physics Division, Oak Ridge National Laboratory, P.O. Box 2008, 
MS-6373,\\ Oak Ridge, TN 37831-6373 USA\\
$^{c}$The Niels Bohr Institute, University of Copenhagen,
Blegdamsvej 17, \\ DK-2100 Copenhagen \O, Denmark}
\maketitle
\begin{abstract}
The properties of the giant-dipole resonance (GDR) 
are calculated as a function of excitation energy, angular momentum, and 
the compound nucleus particle decay width in the nuclei
$^{120}$Sn and $^{208}$Pb, and are
compared with recent experimental data.
Differences observed in the behavior of the full-width-at-half-maximum of 
the GDR for 
$^{120}$Sn and $^{208}$Pb are attributed to the fact that shell corrections 
in $^{208}$Pb are stronger than in $^{120}$Sn, and favor the 
spherical shape at low temperatures. 
The effects shell corrections have on both the free energy and the moments of 
inertia are discussed in detail. 
At high temperature, the FWHM in $^{120}$Sn exhibits effects due to the 
evaporation width of the compound nucleus, while these effects are
predicted for $^{208}$Pb.
\end{abstract}
\pacs{PACS number: 24.30.Cz}

\section{Introduction} 
The study of the properties of the giant-dipole resonance (GDR) at 
finite intrinsic excitation energy has been the objective
of many experimental programs during the past decade (see the reviews in 
Ref.~\cite{r:ref1}).
These experiments yield important information 
regarding nuclear motion as a function of temperature. 
In particular, the role played by quantal and thermal 
fluctuations in the damping of the giant vibrations. In this connection, 
one can individualize the following central issues:
(1)~the temperature dependence of the intrinsic 
width~\cite{r:twidth,r:twidth2}; 
(2)~the time scale for thermal fluctuations testing the 
validity of either the 
adiabatic picture~\cite{r:adiab,r:adiab1,r:adiab2} or the occurrence of 
motional narrowing~\cite{r:mn,r:mn_a}; (3)~the existence 
of a limiting temperature for the 
observation of collective motion in 
nuclei~\cite{r:limit,r:limit_exp}; and (4) the
influence of the lifetime of the compound nucleus on the observed width of the 
GDR~\cite{r:Cho95}. Of particular importance to address these issues is a 
systematic and comprehensive comparison between experiment and theory over a 
wide range of temperatures for several nuclei. 

One of the principal experimental techniques for observing the GDR in 
hot nuclei has been compound-nuclear reactions induced in heavy-ion 
collisions~\cite{r:ref1}. For the most part, the wide range of experiments 
performed so far indicate that the full-width-at-half-maximum (FWHM) 
of the GDR strength function increases as a function of temperature as is 
predicted by theories for the GDR in hot nuclei that account for adiabatic, 
large-amplitude thermal fluctuations of the 
nuclear shape~\cite{r:adiab,r:adiab1,r:adiab2}. 
Many of these experiments, however, 
involve slightly different compound systems and are often 
analyzed using different parameters -- most notably the level-density 
parameter. In addition, because of the dynamics of heavy-ion
collisions, the compound system is generally formed at high angular momentum. 
Indeed, those systems corresponding to the highest excitation energy typically 
have the largest angular momentum content.
As such, it is difficult to separate the effects due 
to large-amplitude thermal 
fluctuations of the shape from those due to angular momentum.

Recently, two experimental methods for studying 
the effects of excitation energy and 
angular momentum separately on the GDR have been introduced. 
In experiments involving compound nuclear reactions, 
large arrays of gamma detectors have been used in order to identify GDR photons 
associated with a system of definite angular momentum. With this experimental
setup, the GDR may be studied within an 
angular momentum window that is usually of the order  
10-15 units of angular momentum wide, and centered between 
30-50~$\hbar$~\cite{r:Bra95}.
An alternative technique is to excite 
a target nucleus by inelastic scattering with light particles~\cite{r:Tho95}, 
which, because of the light mass of the projectile, 
yields an excited system with a fairly small angular momentum. By comparing 
data from these experiments with theoretical predictions, 
it is now possible to analyze the GDR in hot nuclei in 
terms of the effects due to thermal fluctuations and angular 
momentum separately.

In an earlier letter~\cite{r:orm96}, we presented the results
of a systematic study of the FWHM for the 
giant-dipole resonance as a function of temperature, 
angular momentum, and intrinsic width for the nuclei $^{120}$Sn and 
$^{208}$Pb in comparison with recent experimental data from inelastic alpha 
scattering~\cite{r:Tho95}. 
In this work, in addition to providing the details of how this study
was carried out, we also expand upon that work by providing a prediction 
for the influence of the evaporation particles on the FWHM in $^{208}$Pb at 
finite temperature. Because of the systematic analysis over a range of 
temperatures and the relatively low angular momentum of 
the emitting system, it is possible to draw conclusions regarding the
separate roles played by shell corrections, angular momentum,  
and the lifetime of the compound nucleus on the 
observed width of the GDR. 

This work is organized in the following manner. In section II, the formalism
for calculating the effects of thermal fluctuations of the nuclear shape 
while projecting angular
momentum is outlined. A model for the GDR utilizing a quantal, rotating 
harmonic oscillator is given in Section III. A description of the shell
corrections to the free energy and moments of inertia is presented in
Section IV, while results and conclusions are given in Sections V and
VI, respectively.

\section{Thermal fluctuations}
The description of the GDR in hot nuclei begins by noting that 
at a finite temperature, $T$, large-amplitude thermal fluctuations 
of the nuclear shape  
play an important role in the observation of nuclear properties. 
Under the assumption that the time scale associated with thermal 
fluctuations is slow compared to the shift in the dipole frequency 
caused by the fluctuations (adiabatic motion), the GDR cross section 
consists of a weighted average 
over all shapes and orientations. Projecting angular momentum, $J$,
the GDR cross section is evaluated via~\cite{r:Alh93,r:jproj}   
\begin{equation}
\sigma(E) = Z_{J}^{-1} \int 
\frac{{\cal D}[\alpha]}{{\cal I}(\beta,\gamma,\theta,\psi)^{3/2}}
 \sigma(\alpha,\omega_{J};E) 
{\rm e}^{-F(T,\alpha,J)/T},
\label{e:avg_j}
\end{equation}
where $E$ is the photon energy, 
${\cal D}[\alpha]=
\beta^4d\beta\sin(3\gamma)d\gamma\sin\theta d\theta d\phi d\psi$ 
is the volume element, with $\alpha$ denoting the deformation paramters
$\beta$ and $\gamma$ and the Euler angles $\phi$, $\theta$, and $\psi$, 
and  
$Z_J=\int {\cal D}[\alpha]/{\cal I}^{3/2}{\rm e}^{-F/T}$.
The factor ${\cal I}(\beta,\gamma,\theta,\psi)$ is given by
\begin{equation}
{\cal I}(\beta,\gamma,\theta,\psi)= I_1\cos^2\psi\sin^2\theta+
I_2\sin^2\psi\sin^2\theta+I_3\cos^2\theta,
\label{e:inert}
\end{equation}
where the $I_k$ represent the deformation-dependent 
principal moments of inertia. 
The free energy is given by
\begin{equation}
F(T,\alpha,J)=F(T,\alpha,\omega_{rot}=0)+
    (J+1/2)^2/2{\cal I}(\beta,\gamma,\theta,\psi),
\end{equation}
where $F(T,\alpha,\omega_{rot}=0)$ is the free energy evaluated in the 
cranking approximation with rotational frequency, $\omega_{rot}$, equal to 
zero. 

In many previous works~\cite{r:adiab,r:adiab1,r:adiab2}, a 
different procedure involving
a fixed rotational frequency method for projecting angular momentum has been
used. In this formalism, Eq.~(\ref{e:avg_j}) would be replaced by
\begin{equation}
\sigma(E) = Z_\omega^{-1} \int 
{\cal D}[\alpha]
 \sigma(\alpha,\omega;E) 
{\rm e}^{-F(T,\alpha,\omega)/T},
\label{e:avg_om}
\end{equation}
where $Z_\omega=\int {\cal D}[\alpha]{\rm e}^{-F/T}$ and 
the free energy is given by 
\begin{equation}
F(T,\alpha,\omega)=F(T,\alpha,\omega_{rot}=0)-
    \frac{1}{2}{\cal I}(\beta,\gamma,\theta,\psi)\omega^2.
\label{e:free_rot}
\end{equation}
In this scheme, the rotational frequency is determined such that 
the average angular momentum of the system is given 
by~\cite{r:Alh93,r:jproj}
\begin{equation}
\langle J \rangle= J+1/2 = 
T\frac{\partial}{\partial\omega} \ln Z_\omega=
Z_\omega^{-1} \omega \int 
{\cal D}[\alpha]{\cal I}(\beta,\gamma,\theta,\psi)
{\rm e}^{-F(T,\alpha,\omega)/T}.
\end{equation}

The primary disadvantages of the fixed rotational frequency
approach are that angular momentum is projected only on average and that
for finite angular momentum the nuclear free energy in Eq.~(\ref{e:free_rot})
exhibits a saddle
point beyond which the system is unstable. This is illustrated in Fig.~1,
where, in the lower panel, 
the free energy for $^{106}$Sn is plotted along the oblate noncollective
and prolate collective axes at a temperature of 2~MeV and 
a rotational frequency of 1.25~MeV, which 
corresponds to an average angular momentum of approximately $55\hbar$. 
The free energies were computed as described in Section IV, and effectively
consist of only the liquid-drop component. In
the upper panel of Fig.~1, the Boltzman weight factor 
$\exp[-(F-F_{eq})/T]$, where $F_{eq}$ is the minimum of the free energy below
the saddle point, is also plotted. From the figure, it is clear
that at high temperature and high angular momentum, the presence of the
saddle point can be a serious drawback, as it is not possible to
perform the thermal averaging. In addition, an
important shape transition occurring at high spin, known as the Jacobi
transition, which is characterized by the sudden evolution from an oblate 
noncollective shape to a prolate collective shape with large deformation,
is absent. The formalism of  Eq.~(\ref{e:avg_j}) was introduced 
in Ref.~\cite{r:Alh93} to account for these deficiencies and to 
permit a description of the GDR at very high
spin. 

In Ref.~\cite{r:jproj}, an additional method, where only the {\it z}-
component of the angular momentum is projected is also presented. In
this case, Eq.~(\ref{e:avg_j}) is modified to
\begin{equation}
\sigma(E) = Z_{J_z}^{-1} \int 
\frac{{\cal D}[\alpha]}{{\cal I}(\beta,\gamma,\theta,\psi)^{1/2}}
 \sigma(\alpha,\omega_{J_z};E) 
{\rm e}^{-F(T,\alpha,J_z)/T},
\label{e:avg_jz}
\end{equation}
where $Z_{J_z}=\int {\cal D}[\alpha]/{\cal I}^{1/2}{\rm e}^{-F/T}$ and 
the free energy is given by 
\begin{equation}
F(T,\alpha,J_z)=F(T,\alpha,\omega_{rot}=0)+
    (J_z)^2/2{\cal I}(\beta,\gamma,\theta,\psi).
\end{equation}
The principle feature of this projection method is to give a better 
overall description than Eq.~(\ref{e:avg_j}) for 
nonscalar observables such as the angular
distribution $a_2$ coefficient, which is defined by
\begin{equation}
\sigma (E,\theta)=\sigma (E)[1+a_2(E)P_2(\cos\theta)],
\end{equation}
where $\theta$ is the angle between the observed gamma-ray and the
polarized spin direction. In heavy-ion fusion experiments, $J\approx J_z$
and lies in a plane perpendicular to the beam direction, 
and $\theta$ is measured relative to the incident beam direction. 
Then, $a_2$ may be written in terms
spherical tensor components $\sigma_\mu$ of the GDR cross section via
\begin{equation}
a_2(E)=\frac{1}{\sigma (E)}\left[ \sigma_0(E) -\frac{1}{2}(\sigma_1(E)+
\sigma_{-1}(E))\right],
\end{equation}
with $\sigma=\sum_\mu \sigma_\mu$.

Here, we have performed calculations at low spin using all three methods
of angular momentum projection, and find that for the FWHM all three
methods give the same value to within a few hundred keV, with
Eqs.~(\ref{e:avg_om}) and (\ref{e:avg_j}) giving the largest and smallest,
respectively. At much higher spins, $J\approx 50\hbar$, however, 
Eqs.~(\ref{e:avg_j}) and (\ref{e:avg_om}) yield very different results
because of the presence of the saddle-point barrier in the fixed
rotational frequency scheme that does not account for the Jacobi
transition, and limits the effect of thermal fluctuations. 
These issues are discussed in further
detail in Ref.~\cite{r:jproj}.

\section{Model for the GDR}
In principle, the most appropriate description of the GDR strength
function in a hot, rotating nucleus would be obtained by performing 
random phase approximation (RPA) calculations for each deformation 
and orientation. 
Because of the large number of points required in performing the thermal
averaging of Eqs.~(\ref{e:avg_j}), (\ref{e:avg_om}), and (\ref{e:avg_jz}), 
however, this procedure is computationally impractical. 
Instead, we make use of the fact that RPA
calculations indicate that the GDR is a strongly collective excitation
that is also rather stable with temperature~\cite{r:rpa}. As such, for all
practical purposes, the GDR may be modeled by a harmonic vibration along
the three principal nuclear axes with frequencies inversely proportional
to the radius of each axis~\cite{r:Nee82}. Variations of this approach
(with both quantal and classical oscillators),  
have been used in the past~\cite{r:adiab,r:adiab1,r:adiab2,r:mn,r:mn_a}, 
and 
for completeness, we describe in detail the model used in this work in 
the present section. 

A harmonic oscillator description of the GDR may be derived from a many-body 
nuclear Hamiltonian $H$ with a pure harmonic-oscillator single-particle 
potential and an isovector dipole-dipole interaction as the only two-body 
term \cite{r:Boh75,r:Nee82}. 
For the general triaxial nucleus, we have
\begin{equation}
H = {\textstyle {1\over 2}} \sum_{k=1}^3\left[ \sum_{i=1}^A \left(P_k^2+
M\bar\omega_k^2X_k^2\right)_i
+\kappa_k\left(\sum_{i=1}^A(\tau_zX_k)_i\right)^2\right],
\label{e:many}
\end{equation}
where $\tau_z$ is the third component of the isospin, the oscillator 
frequencies $\bar\omega_k$ are inversely proportional to the radius along the 
axis $k=1,2,3$, with
\begin{equation}
\omega_0 = \left(\bar\omega_1\bar\omega_2\bar\omega_3\right)^{1/3}\approx
40A^{-1/3}~{\rm MeV},
\end{equation}
($\hbar=1$), and $\kappa_k$ is the dipole-dipole strength, which empirically is 
of the order $3M\omega_k^2/A$.

In Eq.~(\ref{e:many}), it is possible to introduce a canonical 
transformation in which $H$ is split into two parts. The first 
describing the intrinsic nuclear degrees of freedom, while the second 
the collective GDR mode, which may be written as
\begin{equation}
H_D = {\textstyle {1\over 2}}\sum_k\left(p_k^2+E_k^2d_k^2\right),
\label{e:HD}
\end{equation}
where $d_k$ is the giant-dipole operator and $p_k$ is the conjugate momentum. 
Using the Hill-Wheeler convention \cite{r:Hill53}, the GDR 
resonance energies along the three intrinsic axes are \cite{r:Boh75}
\begin{equation}
E_k = E_0\frac{R_0}{R_k}=E_0\exp\left[-\sqrt{\frac{5}{4\pi}}\beta
\cos\left(\gamma+\frac{2\pi k}{3}\right)\right],
\end{equation}
where $E_0\approx 80A^{-1/3}$~MeV is the dipole energy for the spherical 
shape.

If the intrinsic nuclear frame is rotating with angular velocity $\vec\omega$, 
Eq.~(\ref{e:HD}) must be modified to include the coriolis and centrifugal 
forces, becoming
\begin{equation}
H_D = {\textstyle {1\over 2}}\sum_k(p_k^2+E_k^2d_k^2) -\vec\omega\cdot 
(\vec d\times\vec p),
\label{e:HD_rot}
\end{equation}
where $\vec\omega$ may be taken along the $z$-axis in the external,
fixed reference frame, and while projecting angular momentum is taken to
be the saddle-point value 
$\omega_J=(J+1/2)/{\cal I}(\beta,\gamma,\theta,\psi)$, i.e., the
frequency that maximizes the exponential factors in the projection
integral~\cite{r:Alh93,r:jproj}.
In terms of creation and annihilation operators 
$a_k^{\dag}$ and $a_k$, Eq.~(\ref{e:HD_rot}) may be written as
\begin{equation}
H_D = {\textstyle {1\over 2}}\sum_k E_k(a_k^{\dag}a_k + a_k a_k^{\dag})
+\frac{i}{2} \sum_{ijk}\epsilon_{ijk} \omega_i\sqrt{\frac{E_k}{E_j}}
\left[a_j^{\dag}a_k^{\dag} -a_j^{\dag}a_k + a_ja_k^{\dag}-a_ja_k\right].
\end{equation}
Consolidating the notation, we may write $H_D$ as
\begin{equation}
H_D = {\textstyle {1\over 2}} \sum_{jk}\left( A_{jk}a_j^{\dag}a_k + 
A_{jk}^*a_ka_j^{\dag}
+B_{jk}a_j^{\dag}a_k^{\dag} + B_{jk}^*a_ja_k\right)
\label{HDC}
\end{equation}
with
\begin{equation}
B_{jk} = i\epsilon_{ijk}\omega_i\frac{E_j-E_k}{2\sqrt{E_jE_k}}
\end{equation}
and
\begin{equation}
A_{jk}=\cases{E_j,&if $j=k$;\cr
i\omega_i\frac{E_j+E_k}{2\sqrt{E_jE_k}},&if $i\neq j\neq k.$}
\end{equation}

At this point, we note that $H_D$ is only a quadratic function of the 
coordinates, and, therefore, it is possible to introduce a canonical 
transformation
\begin{eqnarray}
O_\nu^{\dag} &=&\sum_k\left(X_k^\nu a_k^{\dag} - Y_k^\nu a_k\right),\\
O_\nu &=&\sum_k\left(X_k^{\nu *} a_k^{\dag} - Y_k^{\nu *} a_k\right),
\end{eqnarray}
such that the Hamiltonian may be written as
\begin{equation}
H_D={\textstyle {1\over 2}} \sum_\nu E_\nu \left(O_\nu^{\dag}O_\nu + 
O_\nu O_\nu^{\dag}\right).
\end{equation}
The eigenenergies and transformation coefficients $X$ and $Y$ are found from 
the $6\times 6$ RPA-like eigenvalue problem
\begin{equation}
\left(\matrix{A&B\cr A^*&B^*\cr}\right)\left(\matrix{X^\nu\cr Y^\nu\cr}\right)
=E_\nu\left(\matrix{1&0\cr 0&-1\cr}\right)
\left(\matrix{X^\nu\cr Y^\nu\cr}\right).
\end{equation}
Note that the eigenvalues $E_\nu$ come in plus-minus pairs, and the
three principal modes of the GDR correspond to the three positive
eigenvalues.

In order to evaluate the GDR photo-absorption cross section, it is necessary 
to calculate matrix elements of $d_j$. In terms of the creation and 
annihilation operators $O_\nu^{\dag}$ and $O_\nu$ we have
\begin{equation}
d_j = \sqrt{\frac{1}{2E_j}}\left(a_j^{\dag}+a_j\right) =
\sqrt{\frac{1}{2E_j}}\sum_\nu\left[\left(X_j^\nu+Y_j^\nu\right)O_\nu
+\left(X_J^{\nu *}+Y_j^{\nu *}\right)O_\nu^{\dag}\right], 
\end{equation}
and hence the matrix element $\langle\nu\vert d_j\vert 0 \rangle$ can be
written as
\begin{equation}
\langle \nu \vert d_j \vert 0\rangle = 
\sqrt{\frac{1}{2E_j}}\left(X_j^{\nu *}+
Y_j^{\nu *}\right).
\label{e:d_j}
\end{equation}
In addition, the transition matrix elements must be evaluated in the 
non-rotating laboratory frame. This is accomplished by first 
transforming the fixed 
laboratory coordinates to the frame rotating about 
the fixed $z$-axis with rotational frequency $\omega$, and then 
into the intrinsic frame defined by the Euler angles. To do this, 
it is necessary to evaluate the matrix elements of the 
spherical tensors $d_\mu$. Here, we write $d_\mu$ in terms of its
spherical components, that is
\begin{equation}
d_\mu = \sum_j g_{\mu j} d_j,
\end{equation}
where the matrix $g_{\mu j}$ is defined by the well known relations
\begin{equation}
d_\mu=\cases{d_3,& if $\mu=0$;\cr
\mp\frac{1}{\sqrt{2}}(d_1\pm d_2),& if $\mu = \pm 1$.}
\end{equation}
The matrix elements in the frame rotating about the $z$-axis become
\begin{eqnarray}
\langle\nu\vert d_\mu'\vert 0\rangle &=& \sum_{\mu'} 
\langle\nu\vert d_{\mu'}\vert 0\rangle 
D_{\mu \mu'}^{(1)}(\Omega) \nonumber \\
 &=& \sum_{\mu', j} g_{\mu' j}\sqrt{\frac{1}{2E_j}}\left(X_j^{\nu *}
+Y_j^{\nu *}\right) D_{\mu \mu'}^{(1)}(\Omega),
\label{e:d_rot}
\end{eqnarray}
where $D_{\mu \mu'}^{(1)}(\Omega)$ is the rotation function for tensors of 
rank 1.

The GDR cross section to be used in Eqs.~(\ref{e:avg_j}) 
is now readily calculable. From Fermi's Golden rule, 
$\sigma(\alpha,\omega;E)$ evaluated in the intrinsic frame for a 
nucleus with $A$ nucleons, $Z$ protons, and $N$ neutrons is
\begin{equation}
\sigma_{int}(\alpha,\omega;E) = \frac{4\pi^2e^2\hbar}{3mc}\frac{2ZN}{A}
\sum_{\mu,\nu} 
|\langle\nu\vert d_\mu\vert 0\rangle |_{\alpha,\omega}^2
E\left[\delta(E-E_\nu(\alpha,\omega)) - \delta(E+E_\nu(\alpha,\omega))\right].
\end{equation}
Noting that 
\begin{equation}
\delta(E-E') = \frac{1}{2\pi}\int_{-\infty}^{\infty}dt
{\rm e}^{i(E-E')t}
\end{equation}
we have
\begin{eqnarray}
\sigma_{int}(\alpha,\omega;E) &=& \frac{4\pi^2e^2\hbar}{3mc}\frac{2ZN}{A}
\sum_{\mu,\nu}\int_{-\infty}^{\infty}dt 
|\langle\nu\vert d_\mu\vert 0\rangle |_{\alpha,\omega}^2
E\left[{\rm e}^{i(E-E_\nu(\alpha,\omega))t} - 
{\rm e}^{i(E+E_\nu(\alpha,\omega))t}
\right],\nonumber \\
&=& \frac{4\pi^2\hbar}{3mc}\frac{2ZN}{A}
\sum_{\mu,\nu}\int_{-\infty}^{\infty}dt E {\rm e}^{iEt}\nonumber \\
& &\left[
\langle 0 \vert d_\mu^{\dag}(0)\vert\nu\rangle_{\alpha,\omega}
\langle \nu\vert d_\mu(t)\vert 0\rangle_{\alpha,\omega}
-\langle 0\vert d_\mu^{\dag}(t)\vert\nu\rangle_{\alpha,\omega}
\langle\nu\vert d_\mu(0)\vert 0\rangle_{\alpha,\omega}
\right],
\label{e:intrinsic}
\end{eqnarray}
where in the Heisenberg picture $d_\mu(t)={\rm e}^{-iHt}d_\mu(0)
{\rm e}^{iHt}$. The fact that the experimental giant-dipole resonance has an 
intrinsic width, $\Gamma_\nu$, can be accounted for in 
Eq.~(\ref{e:intrinsic}) by 
modifying the exponential by ${\rm e}^{(iE-\Gamma_\nu/2)t}$ giving
\begin{eqnarray}
\sigma_{int}(\alpha,\omega;E) &=& \frac{4\pi^2e^2\hbar}{3mc}\frac{2ZN}{A}
\sum_{\mu,\nu} 
|\langle\nu\vert d_\mu\vert0\rangle |_{\alpha,\omega}^2 E\nonumber \\
& &\big[{\rm BW}(E,E_\nu(\alpha,\omega),\Gamma_\nu) - 
{\rm BW}(E,-E_\nu(\alpha,\omega),\Gamma_\nu)\big],
\label{e:intrinsic2}
\end{eqnarray}
where ${\rm BW}(E,E',\Gamma)$ is a Breit-Wigner function
\begin{equation}
{\rm BW}(E,E',\Gamma) = \frac{1}{2\pi}\frac{\Gamma}
{(E-E')^2+\Gamma^2/4}.
\end{equation}
In the limit that $\omega=0$ (i.e. $E_\nu = E_k$), 
Eq.~(\ref{e:intrinsic2}) 
is a sum of three normalized Lorentzians each with a centroid
at $E_\nu' = \sqrt{E_\nu^2+\Gamma_\nu^2/4}$ and width $\Gamma_\nu$, and 
satisfies 100\% of the classical sum rule. For finite
$\omega$, however, Eq.~(\ref{e:intrinsic2}) is a sum of three Lorentzians
with a normalization of the order $E_\nu/E_k$, and does not necessarily 
satisfy 100\% of the classical sum rule.

Lastly, in order to evaluate $\sigma(\alpha,\omega;E)$ in the non-rotating 
laboratory frame, it is necessary to evaluate the matrix elements 
$\langle\nu\vert d_\mu^{lab}\vert 0\rangle$ in 
Eq.~(\ref{e:intrinsic}). These matrix elements may be 
related to those in the intrinsic frame via Eq.~(\ref{e:d_rot}) 
by noting that the 
transformation from the fixed frame to the rotating frame is accomplished by a 
rotation about the $z$-axis by the angle $\omega t$. That is,
\begin{equation}
d_\mu^{lab}(t) = {\rm e}^{i\mu\omega t} d_\mu'(t) =
\sum_{\mu'} {\rm e}^{i\mu\omega t} d_{\mu'}(t) D_{\mu\mu'}^{(1)}(\Omega).
\end{equation}
From Eq.~(\ref{e:intrinsic}), we see that in addition to mixing 
the strengths of the various components, 
the GDR energies are themselves shifted by 
$-\mu\omega$. Therefore, in the laboratory frame, we have
\begin{eqnarray}
\sigma(\alpha,\omega;E) &=& \frac{4\pi^2e^2\hbar}{3mc}\frac{2ZN}{A}
\sum_{\nu,\mu}\sum_{\mu,\mu'} 
\langle 0\vert d_{\mu'}^{\dag}\vert \nu\rangle_{\alpha,\omega}
\langle\nu\vert d_{\mu''}\vert 0\rangle _{\alpha,\omega}D_{\mu\mu'}^{(1)*}
D_{\mu\mu''}^{(1)}\nonumber \\
& &E\big[ 
{\rm BW}(E,E_\nu(\alpha,\omega)-\mu\omega,\Gamma_\nu)
-{\rm BW}(E,-(E_\nu(\alpha,\omega)-\mu\omega),\Gamma_\nu)
\big].
\label{e:lab0}
\end{eqnarray}
\section{Shell Corrections}
Due to the exponential dependence in Eq.~(\ref{e:avg_j}), 
the most important ingredient for the calculation of the GDR
strength function is the nuclear free energy. Here, 
the free energies were computed using the standard 
Nilsson-Strutinsky~\cite{r:Str66} procedure extended to finite 
temperature~\cite{r:Bra81}, namely
\begin{equation}
F=F_{LD}+F_N-F_S=F_{LD}+F_{SHL},
\label{e:NS}
\end{equation}
where $F_{LD}$ is the liquid-drop free energy evaluated with the
parameters of Ref.~\cite{r:Gue88}, and $F_N$ and $F_S$ are the Nilsson
and Strutinsky components comprising the shell correction, $F_{SHL}$, to the 
free energy. In this work, the Nilsson parameters were taken from
Ref.~\cite{r:Nil69}. For the most part, the 
shell corrections for $^{120}$Sn were found to be quite small (a few hundred 
keV at $T\sim$1.25~MeV), and for all practical purposes can be ignored. 
This is primarily due to the fact that the separate proton and neutron
contributions are approximately equal in magnitude, but opposite in
sign, and, hence, essentially cancel. 
This is in sharp contrast to the strong coherence found in $^{208}$Pb,
where, at low temperatures, strong shell corrections ($\sim - 14$~MeV at 
$T=0$~MeV) are found that favor the spherical shape. 

We have also investigated the influence of the pairing interaction, and have 
found that effects due to pairing are significant only for temperatures below 
$\sim$~0.75~MeV, which is a lower temperature 
than for which experiments have been 
performed. In addition, 
Nilsson-Strutinsky calculations that include pairing, indicate that,
for the most part, the effects on the free energy are negligible. 
This is because $^{208}$Pb is a doubly closed-shell nucleus with pairing
gaps equal to zero for the spherical shape, and in 
$^{120}$Sn, as was the case for the free energy without pairing disscussed 
above, the separate proton and neutron contributions tend to cancel, 
leading to a free energy whose deformation dependence is essentially that of 
the liquid drop.

We note that a numerical determination of the
effects of thermal fluctuations in Eq.~(\ref{e:avg_j})  
in general requires an exploration of the five dimensional 
space spanned by the deformation and orientation degrees of freedom, in which a
large number of points are required in order to assure sufficient accuracy
(especially at finite angular momentum). In
this regard, a Nilsson-Strutinsky calculation for each point may be too time
consuming. Therefore, it is useful to parameterize the free energy using
functions that mimic the behavior of the Nilsson-Strutinsky calculation as
closely as possible. It has been pointed out~\cite{r:Lev84} that, being a
scalar quantity, the free energy must be a function of the rotational
invariants of the quadrupole deformation, that is 
\begin{equation}
F(T,\beta,\gamma) = F_0(T) +A(T)\beta^2-B(T)\beta^3\cos(3\gamma)+
C(T)\beta^4 +...
\label{e:landau}
\end{equation}
Although this Landau parameterization gives a good overall description of 
the free energy, in particular regarding to shape transitions, it may not 
be adequate for the evaluation of Eq.~(\ref{e:avg_j}) because at somewhat 
larger deformations Eq.~(\ref{e:landau}) deviates from the Nilsson-Strutinsky 
calculation,
often giving a much stiffer free energy. This is principally because
Eq.~(\ref{e:landau}) attempts to combine both the liquid-drop free 
energy and shell
corrections, $F_{SHL}=F_N-F_S$, into the same parameterization. An alternative 
approach is to 
parameterize instead only the shell corrections to the free energy using a
function of the rotational invariants. 

Exhibited in Fig. 2 (solid points) are shell corrections to the 
free energy at $\omega_{rot}=0$ as a function of temperature for
oblate ($\gamma=\pi/3$), prolate ($\gamma=0$), and triaxial ($\gamma=\pi/6$) 
deformations for $^{208}$Pb. The general
overall behavior of $F_{SHL}$ is to decrease with temperature, gradually
melting ($F_{SHL}\approx 0$~MeV) for temperatures of the order $T=2.5$~MeV, and
that they tend to oscillate with deformation, but appearing to be damped at
larger $\beta$. In this light, it is possible to
parameterize $F_{SHL}$ with a series of functions that are in fact 
themselves functions of the rotational invariants $\beta^2$, 
$\beta^3\cos(3\gamma)$, etc... One possible parameterization is 
\begin{eqnarray}
F_{SHL}(\beta,\gamma,T) &=&\sum_{l=0}^{even} A_l j_l(B_l\beta)C_lT
/{\rm sinh}(C_lT)\nonumber\\
&+&\sum_{l=3}^{odd} A_l j_l(B_l\beta)\cos(3\gamma)C_lT/{\rm sinh}(C_lT),
\label{e:free_bess}
\end{eqnarray}
where the $j_l$ are spherical Bessel functions. We note that 
$C_lT/{\rm sinh}(C_lT)$ is the expected attenuation behavior as a 
function temperature when the 
single-particle Hamiltonian is a 
degenerate harmonic oscillator~\cite{r:Boh75}.

The parameters $A_l$, $B_l$, and $C_l$ can be determined in the following 
manner. 
First, carry out a Nilsson-Strutinsky calculation for oblate, prolate and 
triaxial shapes up to $\beta=1.0$, and for temperatures between $T=0.25$ and 
3.0~Mev.
Then fit both parameters $A_l$ and $B_l$ to the Nilsson-Strutinsky 
calculation at $T=0.25$~MeV for all three shapes simultaneously. 
Note that at
$\beta=0$, the free energy is completely determined by $A_0$, and as
such is not fit upon. In addition, note that the $\gamma=\pi/6$ points
are dependent only on the even functions in Eq.~(\ref{e:free_bess}).
Typically, the number of terms in 
Eq.~(\ref{e:free_bess}) can be truncated to $l\sim 5$.
With the parameters $B_l$ determined at $T=0.25$~MeV, these 
same values are then used to fit the $A_l$ values at all other temperatures, 
giving the sequence $\{A_l(T_i)\}$, which is then
fit to the function $A_lC_lT/{\rm sinh}(C_lT)$. Shown
in Fig. 2 with the solid line are the results of the parameterization of
the shell corrections to the free energy for $^{208}$Pb, and the associated
parameters are listed in Table I. Generally speaking, the parameterization
of Eq.~(\ref{e:free_bess}) gives a good overall reproduction of the shell
corrections, $F_{SHL}$, that is rather quick to implement with 
Eq.~(\ref{e:avg_j}). 

We note one feature of the parameterization for $^{208}$Pb is that
the parameterized shell corrections tend to ``melt'' a little too quickly. 
For example, for $T > 1.5$~MeV the parameterized shell corrections 
underestimate the
Nilsson-Strutinsky values by a few hundred keV. It is to be noted, however,
that at these temperatures, 
this amounts to a relatively small change in the overall 
deformation dependence of the total free energy, which, in addition to
be divided by the temperature in the Boltzman factor, ${\rm e}^{-F/T}$ is  
at basically dominated by the liquid-drop component. 

In addition to the free energy, shell structure can also modify the
moments of inertia. Again, we employ the Nilsson-Strutinsky procedure at
finite rotational frequency, and obtain shell corrections to the rigid-body 
moments of inertia, namely
\begin{equation}
I=I_{rigid}+I_N-I_S=I_{rigid}+I_{SHL},
\end{equation}
where here the rigid-body values were evaluated with the radius
$R=1.2A^{1/3}$.
Choosing the rotational frequency along the 
$z$-axis, the leading behavior as a function of
rotational frequency for each of the free energy 
components in Eq.~(\ref{e:NS}) is given by
\begin{equation}
F(\beta,\gamma,T,\omega)=F(\beta,\gamma,T,\omega=0)-{\textstyle {1\over2}}
I_3(\beta,\gamma,T)\omega^2.
\end{equation}
The moments of inertia $I_3$ can then be obtained by performing a
quadratic fit to the free energy components.

In a manner similar to the shell corrections to the free energies, the shell 
corrections to the moments of inertia may also be parameterized by
series of Bessel functions, i.e., 
\begin{eqnarray}
I_3^{SHL}(\beta,\gamma,T) &=&\sum_{l=0}^{even} A_l^I j_l(B_l^I\beta)C_l^IT
/{\rm sinh}(C_l^IT)\nonumber\\
&+&\sum_{l=3}^{odd} A_l^I j_l(B_l^I\beta)\cos(3\gamma)C_l^IT/
{\rm sinh}(C_l^IT)\nonumber \\
&+&\sum_{l\geq 1}\alpha_lj_l(\kappa_l\beta)\cos(\gamma+2\pi/3)\eta_lT/
{\rm sinh}(\eta_lT),
\label{e:inert_bess}
\end{eqnarray}
where the third term in the sum is included because of rotational
invariance arguments for the moment of inertia~\cite{r:adiab1}. Once the
third component of the moment of inertia is determined as a function of
$T,\beta,\gamma$, the remaining two components are obtained by the
relations~\cite{r:adiab1}
\begin{eqnarray}
& &I_1(T,\beta,\gamma)=I_3(T,\beta,\gamma+2\pi/3),\nonumber\\
& &I_2(T,\beta,\gamma)=I_3(T,\beta,\gamma-2\pi/3).
\end{eqnarray}

The parameters $A_l^I$, $B_l^I$, $C_l^I$, $\alpha_l$, $\kappa_l$ and $\eta_l$
were determined in a similar manner 
as those for the free energy 
in Eq.~(\ref{e:free_bess}). Again, the shell corrections for $^{120}$Sn 
were found to be
small and negligible. For comparison, both the parameterized and 
Nilsson-Strutinsky shell corrections to the moment of inertia for
$^{208}$Pb are shown in Fig.~3 as a function of temperature and for the
deformations $\gamma=\pi /3,~\pi/6,~0,~-\pi/3,~{\rm and}~-2\pi/3$. The
most important feature is the strong shell corrections at $\beta=0$ that
significantly reduced the moment inertia below the rigid-body value.

Of particular importantance for the moments of inertia is the fact that
the spin-orbit and $l^2$ terms in the Nilsson Hamiltonian lead to
moments of inertia that are approximately 20-30\% larger than the
corresponding rigid-body values~\cite{r:Nee77}. As such, the shell
corrections to the moments of inertia used here 
were reduced by 25\%, which 
corresponds to the average difference between the rigid-body and Strutinsky
moments of inertia. The corresponding parameters  
$A_l^I$, $B_l^I$, $C_l^I$, $\alpha_l$, $\kappa_l$ and $\eta_l$ 
are then listed in Table II for $^{208}$Pb.

Because of the ${\cal I}^{-3/2}$ dependence in the ``effective'' volume
element in Eq.~(\ref{e:avg_j}), it might be expected that the strong
shell corrections to the moment of inertia in $^{208}$Pb would
significantly affect the GDR strength function, as they  
appear to give a stronger preference to the spherical shape. 
We find, however, that
because of the $\beta^4$ factor in ${\cal D}[\alpha]$, the strong shell
corrections in ${\cal I}$ 
favoring the spherical shape have very little effect on the 
FWHM of the GDR at low spin beyond that produced by 
the free energy. This is exhibited in Fig.~4, where an ``effective''
weight factor (which for the sake of simplicity ignores the $\sin3\gamma$
factor)
$W=\beta^4/{\cal I}^{3/2}{\rm e}^{-F/T}$ is plotted for 
oblate and prolate shapes at $T=1.0$~MeV for various combinations of
the shell corrections. In the top panel of the figure, the weight factor
is plotted including shell corrections to the free energy as well as
with and without shell corrections to the moments of inertia, whereas
the corresponding figure without shell corrections to the free energy is
shown in the bottom part of the figure. In both cases, it is seen that
the overall behavior of the weight function is governed by the exponential 
of the free energy, which is plotted in the upper right-hand panel. In 
addition, the ratio ${\cal I}_{LD}/{\cal I}_{SHL}$ is shown in the lower
left-hand panel, where it is seen that without the $\beta^4$ factor the
spherical shape would have approximately 40\% more weight when shell 
corrections to the moments of inertia are included.

\section{Results}
In this section we present the results of a systematic comparison 
between theoretical calculations and recent experimental data~\cite{r:Tho95} 
as a function of temperature for both $^{120}$Sn and $^{208}$Pb. The thermally
averaged GDR cross sections were computed using Eqs.(\ref{e:avg_j}) and
(\ref{e:lab0}). In keeping with experimental findings~\cite{r:Car74},
the intrinsic dipole widths, $\Gamma_\nu$ were taken to be dependent on
the centroid energy $E_\nu$ via $\Gamma_\nu=\Gamma_0(E_\nu/E_0)^\delta$, 
where $E_0$ and $\Gamma_0$ are the centroid and width for spherical shape and
$\delta\approx 1.8$. The parameters $E_0$ and $\Gamma_0$ were 
taken from ground-state data and are  
$E_0=14.99$~MeV and $\Gamma_0=5.0$~MeV for $^{120}$Sn and $E_0=13.65$~MeV and 
$\Gamma_0=4.0$~MeV for $^{208}$Pb, respectively. Finally, in accordance
with the considerable theoretical evidence presented in Ref.~\cite{r:twidth}, 
the intrinsic width $\Gamma_0$ is taken to be independent
of temperature throughout this work.

Shown in Fig.~5 are the results obtained for the FWHM of the GDR strength
function for both $^{120}$Sn and $^{208}$Pb as a function of temperature in 
comparison with recent experimental data. The 
solid line represents the theoretical values obtained with zero angular 
momentum. The dependence of the FWHM for $^{120}$Sn and $^{208}$Pb 
on angular momentum
at $T=1.6$~MeV is illustrated 
in Fig.~6, where it is seen that for $J\leq 25\hbar$ the FWHM is 
essentially unchanged from the $J=0\hbar$ value. 
Given that the largest average 
angular momentum in the systems studied experimentally is of the 
order $20\hbar$~\cite{r:Tho95}, the effects due to 
angular momentum on the data set of interest are then 
expected to be negligible.

As is seen from Fig.~5, theory provides an overall account of the 
experimental findings. The dependence of the FWHM on temperature 
is quite different between $^{120}$Sn and $^{208}$Pb.
The FWHM in $^{208}$Pb appears to be suppressed at lower 
temperatures relative to $^{120}$Sn. This is due to the rather strong 
shell corrections in 
$^{208}$Pb that favor the spherical shape at low temperatures. 
The affect of such strong 
shell corrections is to limit the influence of thermal fluctuations at low 
temperatures, thereby reducing the observed width. This is also illustrated
in Fig.~5, where the dotted line in the panel for $^{208}$Pb indicates the FWHM 
assuming no shell corrections. 
We note that the shell correction effect and the angular momentum dependence 
was also observed for $^{140}$Ce in 
Ref.~\cite{r:adiab2}. 
The fact that the adiabatic model slightly overestimates the FWHM maybe due to: 
(1) uncertainties in the extracted temperature; (2) the 
shell corrections being more persistent at higher temperatures 
than predicted by Nilsson-Strutinsky calculations; 
(3) the fact that the experimental strength functions 
were fit to a single Lorentzian, while theoretically they are 
obtained from the superposition of many Lorentzians; 
and/or (4) the presence of non-adiabatic effects that would lead to 
a motional narrowing of the FWHM~\cite{r:mn}. 
In keeping with point (1) above, one can mention that 
the temperatures inferred from experiment are 
sensitive to the choice of the level-density parameter, and, as a 
consequence, are uncertain at the level of approximately 0.2~MeV. 

The FWHM shown in Fig. 5 are essentially consistent with the adiabatic picture 
for the GDR in hot nuclei, and do not present any evidence for 
the phenomenon known as motional narrowing~\cite{r:mn,r:mn_a}, which 
tends to lessen the effects of thermal broadening on the resonance, and, 
hence, reduce the FWHM. As is pointed out in 
Ref.~\cite{r:mn_a}, however, because of a lack of reliable theoretical 
estimates for the time scales associated with thermal fluctuations, 
the FWHM is not sufficient in of itself to exclude 
motional narrowing. This is particularly true when the time scales 
for $\beta$ and $\gamma$ degrees of freedom are much faster than
those associated with the orientation of the system. In this case, both 
the response function and the angular distribution $a_2$ coefficient are 
needed in order to make a differentiation between the two regimes.

We note some slight discrepancies between the adiabatic model 
and experiment for $^{120}$Sn. 
To begin with, the FWHM at $T=1.24$~MeV is significantly lower than the 
theoretical prediction and is difficult to explain within the framework of the 
model. This datum seems to point to the existence of strong 
shell corrections that quickly disappear at $T=1.5$~MeV, which is in 
disagreement with the expectations of the Nilsson-Strutinsky procedure.
At higher temperatures, $T\approx 2.8-3.1$~MeV, the experimental FWHM is 
somewhat larger 
than the theoretical values, and may indicate a systematic trend to be observed 
at still higher temperatures. Shown in Fig.~7 is the FWHM for $^{120}$Sn at 
$T=3.12$~MeV as a function of the intrinsic width $\Gamma_0$. At this 
temperature, the experimental FWHM is $11.5\pm 1.0$~MeV, and we may infer from 
this datum a value of $\Gamma_0=7.7^{+1.8}_{-2.1}$~MeV, as indicated by the 
solid square (11.5~MeV) and open circles ($\pm 1$~MeV) in Fig.~7.
We note, however, that this is consistent with 
the concept that the width observed for the GDR in hot nuclei should be 
increased because of the evaporation of particles from the 
compound nucleus~\cite{r:Cho95}.
At higher excitation energies, the decay rate for particle evaporation 
increases, and, because of the uncertainty principle, the energy of an emitted 
GDR photon cannot be known with a precision better than 
$\Gamma_{cn}=\Gamma_{ev}^{before} + \Gamma_{ev}^{after}$, where 
$\Gamma_{ev}^{before(after)}$ is the width 
for particle evaporation {\it before} and {\it after} the 
emission of the GDR photon. To account for this effect in our calculations,
we note that the 
FWHM shown in Fig.~5 are obtained from the full response function, which also 
includes splittings due to the superposition of the various intrinsic modes. 
On the other hand, $\Gamma_{cn}$ represents an uncertainty in the GDR photon 
energy due to the lifetime of the initial and final states.
As such, $\Gamma_{cn}$ should be folded into the GDR response function 
by increasing the intrinsic widths via 
$\Gamma_\nu^\prime \rightarrow \Gamma_\nu+\Gamma_{cn}$.
In order to estimate $\Gamma_{cn}$ for $^{120}$Sn we refer to Fig.~2 of  
ref.~\cite{r:limit}, where $\Gamma_{ev}$ is plotted as a function of 
excitation energy for various values of the level-density parameter $a$
(which is conventionally defined as $a=A/\kappa$~MeV$^{-1}$, and values of 
$\Gamma_{ev}$ are plotted for $\kappa=8$,~10, and 12). We note that at a 
given excitation energy, $\Gamma_{ev}$ exhibits a strong
dependence on $a$. Indeed, at $E_x=150$~MeV, there is a nearly a factor
of three difference between the results for 
$\kappa=8$ and 12. This rather strong dependence
on $a$ is considerably diminished, however, when converting to temperature,
defined as $E_x-E_{GDR}=aT^2$, as is shown in Table III where 
$\Gamma_{ev}^{before(after)}$ and $\Gamma_{cn}$
are given as a function of temperature for $\kappa=10$ and 12. Only at the
highest temperatures ($\approx 3.5$~MeV), where $\kappa$ is expected to 
be of the order 12-13, is the difference much greater than a few hundred
keV. Taking $\kappa=12$, we deduce at $T\approx 3.1$~MeV 
$\Gamma_{cn}\approx 2.1$~MeV, which is in good 
agreement with the experimental results as is illustrated in Fig.~7. 
To further see the influence of the evaporation width, we have 
computed the FWHM for $^{120}$Sn a function of temperature 
including $\Gamma_{cn}$ (evaluated with $\kappa=12$), which is shown in Fig.~5
by the dashed line. On the whole, the inclusion of $\Gamma_{cn}$ 
leads to a better overall agreement with experiment.

It is to be noted that although the experimental data for $^{208}$Pb do
not, as yet, extend to $T\sim 3.0$~MeV, the effects of the particle evaporation 
width will also be present in $^{208}$Pb. We have computed $\Gamma_{ev}$
for $^{208}$Pb using the same method as in Ref.~\cite{r:limit} and is 
displayed in Fig.~8 as a function of excitation energy for $\kappa=8$,~10, 
and 12. Also shown in Table IV are values of $\Gamma_{ev}^{before(after)}$
and $\Gamma_{cn}$ as a function of temperature for $\kappa=10$ and 12.
The FWHM for $^{208}$Pb including $\Gamma_{cn}$ (with $\kappa=12$) is 
shown in Fig.~5 with the dashed line, where it is seen that at
$T=3.25$~MeV the FWHM is approximately 2.5~MeV larger than predicted by the
adiabatic model. As such, experiments carried out in this temperature
range would be a further confirmation of this effect. Finally, as is
pointed out in Ref.~\cite{r:limit}, the particle evaporation width also
leads to a maximum excitation energy (or limiting temperature) above 
which the GDR is not observable. This occurs when $\Gamma_{ev}\sim\Gamma_0$, 
which for $^{208}$Pb corresponds to 
$E_x\approx 300-350$~MeV (or $T\approx 4.2-4.5$~MeV).

\section{conclusions}
We conclude that a systematic study of the FWHM of the GDR as a 
function of temperature for the nuclei $^{120}$Sn and $^{208}$Pb confirms,  
for the first time, the 
overall theoretical picture of the GDR in hot nuclei at low spin. 
In particular, the role 
played by adiabatic, large-amplitude thermal fluctuations of the nuclear shape. 
In fact, overall agreement between theory and experiment is observed 
over a range of temperatures for both $^{120}$Sn and $^{208}$Pb, 
which display quite 
different behaviors for the FWHM as a function of temperature. 
This difference can be 
attributed to the presence of strong shell corrections 
favoring spherical shapes in $^{208}$Pb that are absent in $^{120}$Sn. 
Finally, the increase in the FWHM over that expected from thermal averaging 
at temperatures of the order 3.0~MeV is 
in accordance with the increase expected from the particle evaporation of the 
compound system.

\begin{center}
{\bf ACKNOWLEDGEMENTS}
\end{center}
Oak Ridge National Laboratory is managed for the U.S. Department of Energy
by Lockheed Martin Energy Research Corp. under contract No.
DE--AC05--96OR22464. 
\newpage

\bibliographystyle{try}

\begin{table}
\caption{Parameters in Eq.~(38) to define the shell
corrections to the free energy in $^{208}$Pb.}
\begin{tabular}{cccc}
$l$ & $A_l$ & $B_l$ & $C_l$ \\
\tableline
  0 & -13.706 & 13.764 & 3.011 \\
  2 & -6.448  & 10.357 & 3.122 \\
  3 & 6.68 & 8.159 & 2.408 \\
  5 & 19.50 & 23.882 & 3.146 \\
\end{tabular}
\end{table}

\begin{table}
\caption{Parameters in Eq.~(41) to define the shell
corrections to the moment of inertia in $^{208}$Pb.}
\begin{tabular}{cccc}
$l$ & $A_l$ & $B_l$ & $C_l$ \\
\tableline
  0 & -87.653 & 13.764 & 3.108 \\
  2 & -49.343  & 10.357 & 3.026 \\
  3 & 28.532 & 8.159 & 3.068 \\
  5 & 80.810 & 23.791 & 3.558 \\
\tableline
$l$ & $\alpha_l$ & $\kappa_l$ & $\eta_l$ \\
  1 & -22.910 & 8.334 & 2.979 \\
\end{tabular}
\end{table}

\begin{table}
\caption{Values of 
$\Gamma_{cn}=\Gamma_{ev}^{before}+\Gamma_{ev}^{after}$ 
for $^{120}$Sn 
obtained from Fig. 2 of Ref.~[10] as a function of temperature
and the level-density parameter defined as $a=A/\kappa$~MeV$^{-1}$. 
All quantities are given in MeV}
\begin{tabular}{cccccccccc}
 $T$ &  & $\kappa=10$ &  &  &  &  & $\kappa=12$ &  &  \\
  & $E_x$ & $\Gamma_{ev}^{before}$ & $\Gamma_{ev}^{after}$ &
 $\Gamma_{cn}$ &  & $E_x$ & $\Gamma_{ev}^{before}$ & $\Gamma_{ev}^{after}$ &
 $\Gamma_{cn}$\\
\tableline
 1.25 & 34 & 0.03 & 0.00 & 0.03 & & 31 & 0.06 & 0.02 & 0.08 \\
 1.50 & 42 & 0.06 & 0.03 & 0.09 & & 38 & 0.10 & 0.05 & 0.15 \\
 1.75 & 52 & 0.09 & 0.06 & 0.17 & & 46 & 0.13 & 0.06 & 0.19 \\
 2.00 & 63 & 0.14 & 0.08 & 0.22 & & 55 & 0.20 & 0.11 & 0.31 \\
 2.25 & 76 & 0.22 & 0.12 & 0.34 & & 66 & 0.33 & 0.17 & 0.50 \\
 2.50 & 90 & 0.37 & 0.21 & 0.58 & & 78 & 0.52 & 0.29 & 0.81 \\
 2.75 & 106 & 0.61 & 0.38 & 0.99 & & 91 & 0.71 & 0.49 & 1.20 \\
 3.00 & 123 & 0.83 & 0.63 & 1.46 & & 105 & 0.95 & 0.70 & 1.65 \\
 3.25 & 142 & 1.12 & 0.89 & 2.01 & & 121 & 1.31 & 0.97 & 2.28 \\
 3.50 & 162 & 1.45 & 1.19 & 2.64 & & 138 & 1.72 & 1.36 & 3.08 \\
\end{tabular}
\end{table}

\begin{table}
\caption{Values of 
$\Gamma_{cn}=\Gamma_{ev}^{before}+\Gamma_{ev}^{after}$ 
for $^{208}$Pb as a function of temperature
and the level-density parameter defined as $a=A/\kappa$~MeV$^{-1}$. 
All qunatities are given
in MeV.}
\begin{tabular}{cccccccccc}
 $T$ &  & $\kappa=10$ &  &  &  &  & $\kappa=12$ &  &  \\
  & $E_x$ & $\Gamma_{ev}^{before}$ & $\Gamma_{ev}^{after}$ &
 $\Gamma_{cn}$ &  & $E_x$ & $\Gamma_{ev}^{before}$ & $\Gamma_{ev}^{after}$ &
 $\Gamma_{cn}$\\
\tableline
 1.50 & 61 & 0.05 & 0.02 & 0.07 & & 53 & 0.07 & 0.02 & 0.09 \\
 1.75 & 77 & 0.10 & 0.04 & 0.14 & & 67 & 0.11 & 0.06 & 0.17 \\
 2.00 & 97 & 0.24 & 0.12 & 0.36 & & 83 & 0.23 & 0.13 & 0.36 \\
 2.25 & 119 & 0.39 & 0.27 & 0.66 & & 102 & 0.46 & 0.27 & 0.73 \\
 2.50 & 144 & 0.64 & 0.47 & 1.13 & & 122 & 0.65 & 0.50 & 1.15 \\
 2.75 & 171 & 0.90 & 0.76 & 1.66 & & 145 & 1.01 & 0.77 & 1.78 \\
 3.00 & 201 & 1.33 & 1.10 & 2.43 & & 170 & 1.38 & 1.15 & 2.53 \\
 3.25 & 234 & 1.73 & 1.52 & 3.25 & & 197 & 1.90 & 1.58 & 3.48 \\
 3.50 & 269 & 2.20 & 2.01 & 4.21 & & 226 & 2.42 & 2.12 & 4.54 \\
\end{tabular}
\end{table}

\begin{figure}
\caption{The free energy for $^{106}$Sn is plotted (lower panel)
along the oblate noncollective ($\beta < 0$)
and prolate collective ($\beta > 0$) axes at a temperature of 2~MeV and 
a rotational frequency of 1.25~MeV $(\langle J\rangle\approx 55\hbar)$. In
the upper panel, the Boltzman weight factor 
$\exp[-(F-F_{eq})/T]$, where $F_{eq}$ is the minimum of the free energy below
the saddle point, is plotted.}
\end{figure}

\begin{figure}
\caption{Nilsson-Strutinsky shell corrections (solid squares) 
to the free energy for 
$^{208}$Pb as a function of temperature for oblate ($\gamma=\pi/3$),
triaxial ($\gamma=\pi/6$), and prolate ($\gamma=0$) shapes at zero
angular momentum.
The parameterization to the shell corrections
given by Eq.~(38) is represented by the solid line.}
\end{figure}

\begin{figure}
\caption{Nilsson-Strutinsky shell corrections (solid squares) to the
moments of inertia for $^{208}$Pb for $\gamma=\pi/3$, $\pi/6$, 0, $-2\pi/3$,
and $-\pi/3$. The parameterization to the shell corrections
given by Eq.~(41) is represented by the solid line.
The temperature for each panel is the same as in Fig.~2.}
\end{figure}

\begin{figure}
\caption{The weight function 
$W(\beta)=\beta^4/{\cal I}^{3/2}{\rm e}^{-F/T}$ at $T=1.0$~MeV
for prolate ($\beta >0$) and oblate ($\beta < 0$) shapes. In panel (a), 
$W(\beta)$ includes shell corrections to the free energy,
$F_{SHL}$, as well as with (dotted line) and without (solid line) 
shell corrections to the moments of inertia. In panel~(b), 
the same quantities are plotted without shell corrections to the free
energy, i.e., $F=F_{LD}$. The free energy with and without shell
corrections is plotted in panel (c), and  
the factor $({\cal I}_{LD}/{\cal I}_{SHL})^{3/2}$ is plotted in 
panel~(d).}
\end{figure}

\begin{figure}
\caption{The FWHM of the GDR strength function as a function 
of temperature for $^{120}$Sn and $^{208}$Pb. Experimental data are represented 
by the filled circles, while the solid line represents the theoretical results 
obtained for $J=0\hbar$. For $^{208}$Pb, 
the dotted line is the FWHM obtained assuming no shell corrections.
For $^{120}$Sn and $^{208}$Pb, the dashed line represents the FWHM 
obtained by including the increase to the intrinsic width, $\Gamma_{cn}$, 
due to the evaporation of particles from the compound system.}
\end{figure}

\begin{figure}
\caption{The FWHM in $^{120}$Sn (dashed line) and $^{208}$Pb (solid line) 
at $T=1.6$~MeV as a function of angular momentum.}
\end{figure}

\begin{figure}
\caption{The FWHM in $^{120}$Sn at $T=3.12$~MeV as a function of the intrinsic
width $\Gamma_0$ (solid line). The experimental value of $11.5\pm 1.0$~MeV is 
represented by the filled square (11.5~MeV) and the open circles 
($\pm 1$~MeV).}
\end{figure}

\begin{figure}
\caption{The particle evaporation width, $\Gamma_{ev}$, for $^{208}$Pb 
as a function of excitation energy for three values of the level-density
parameter, i.e. $\kappa=8$, 10, and 12 (note $a=A/\kappa$~MeV$^{-1}$).}
\end{figure}

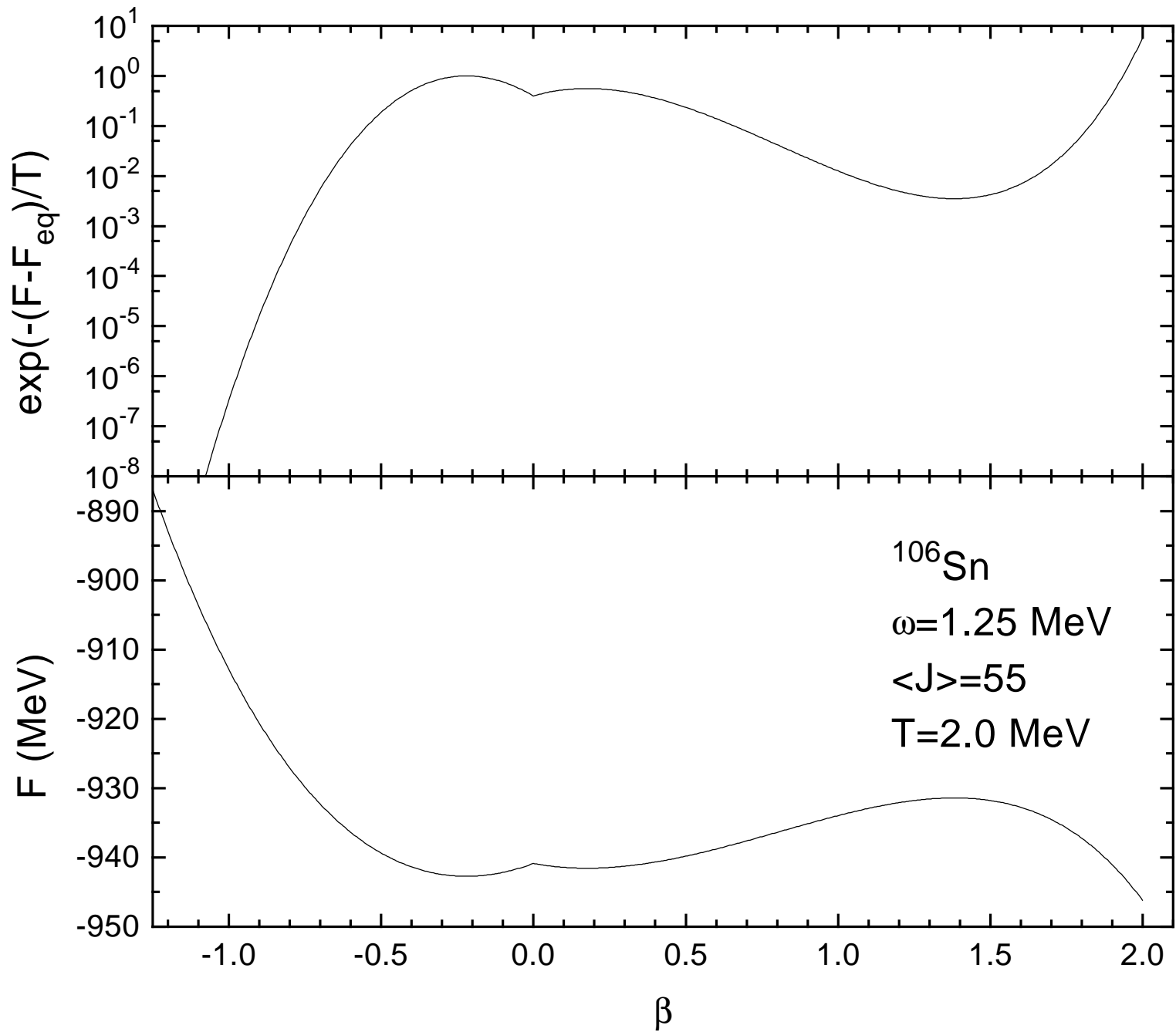
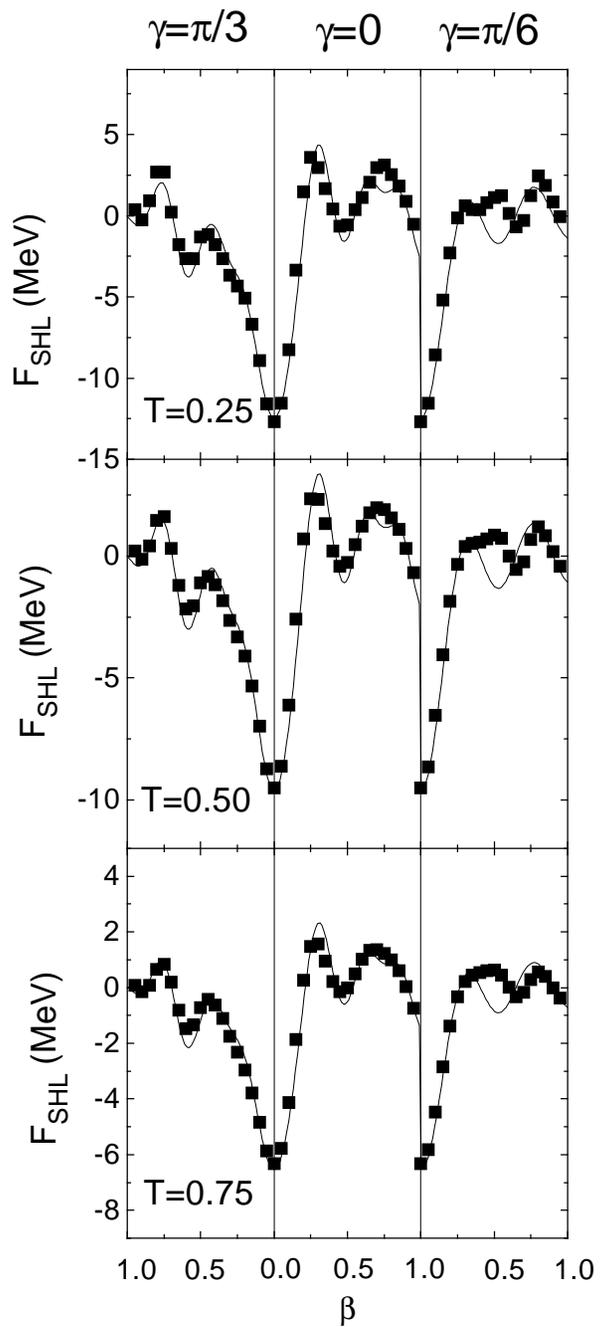
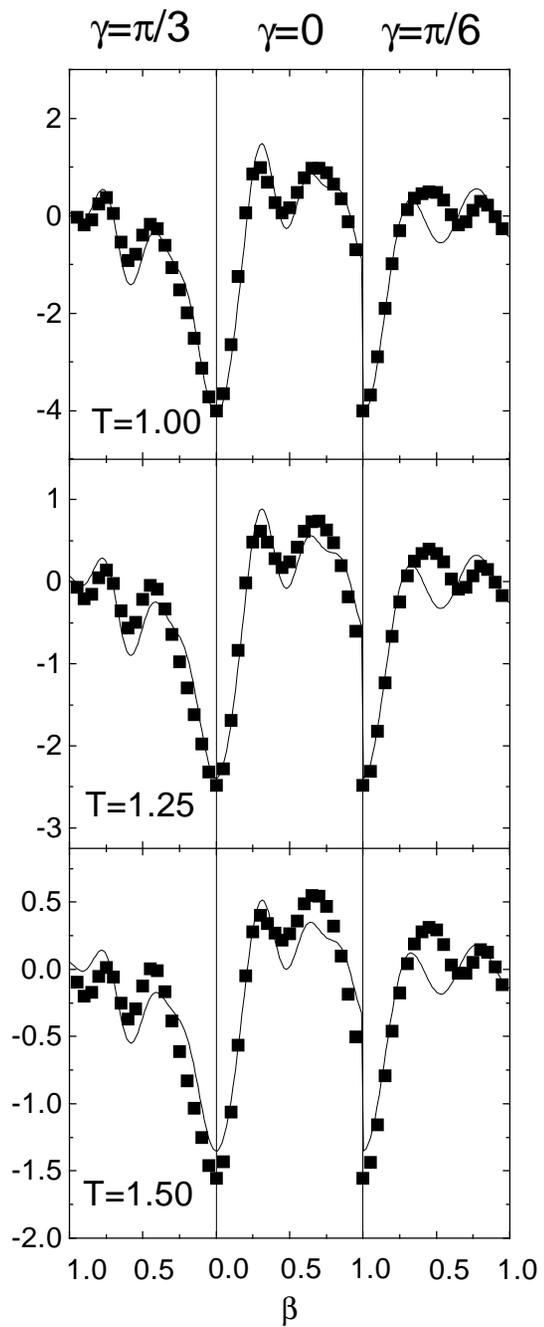
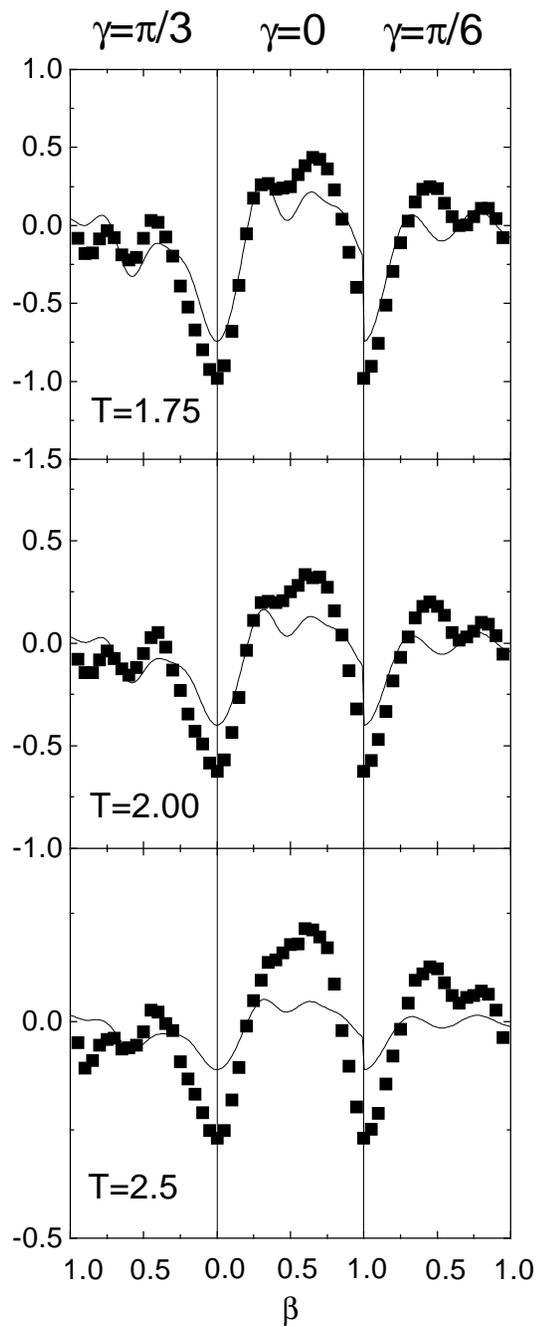
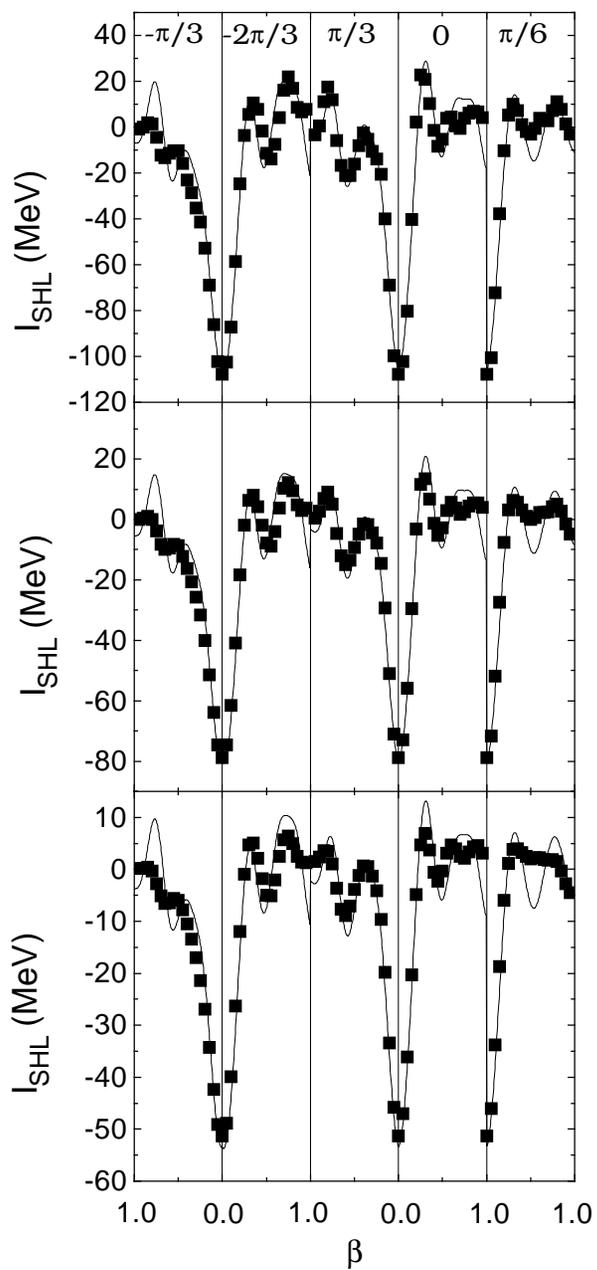
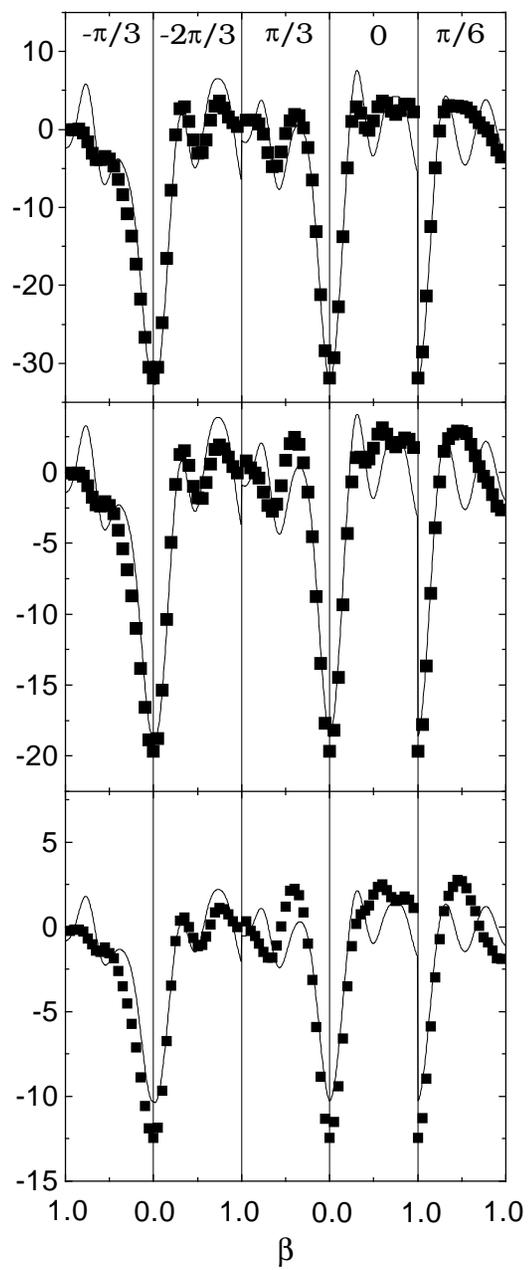
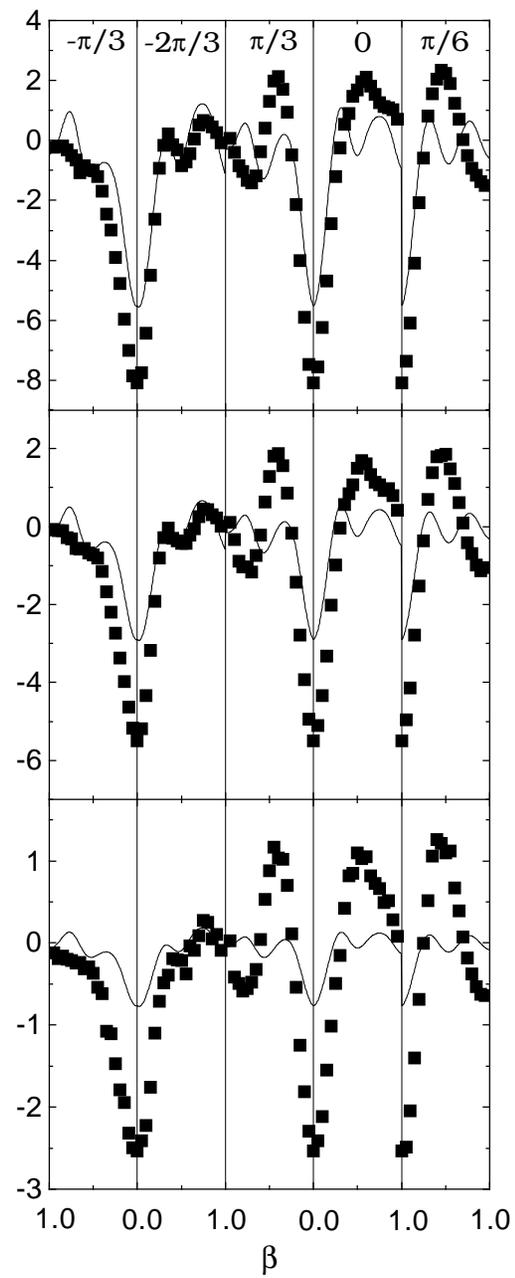
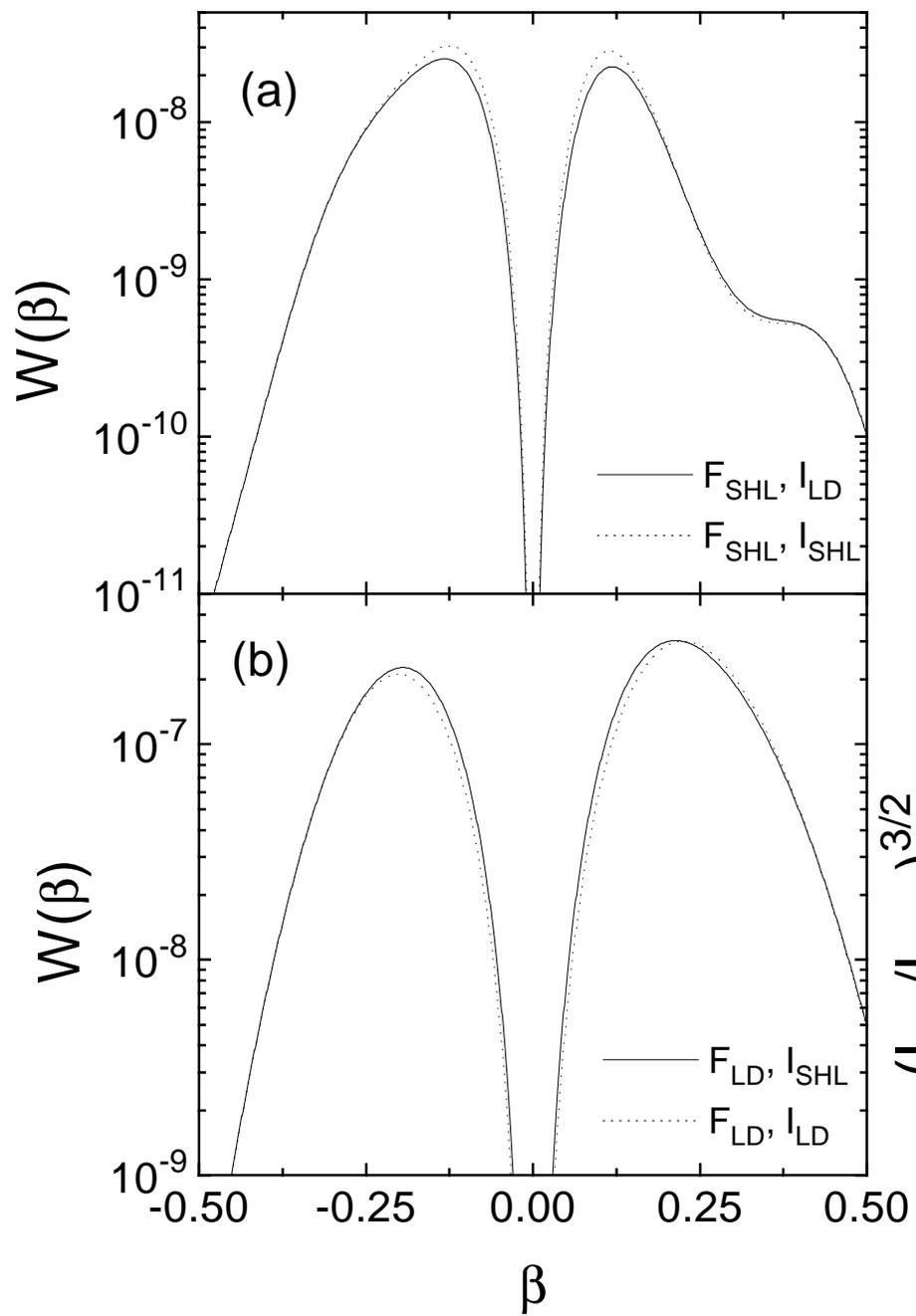
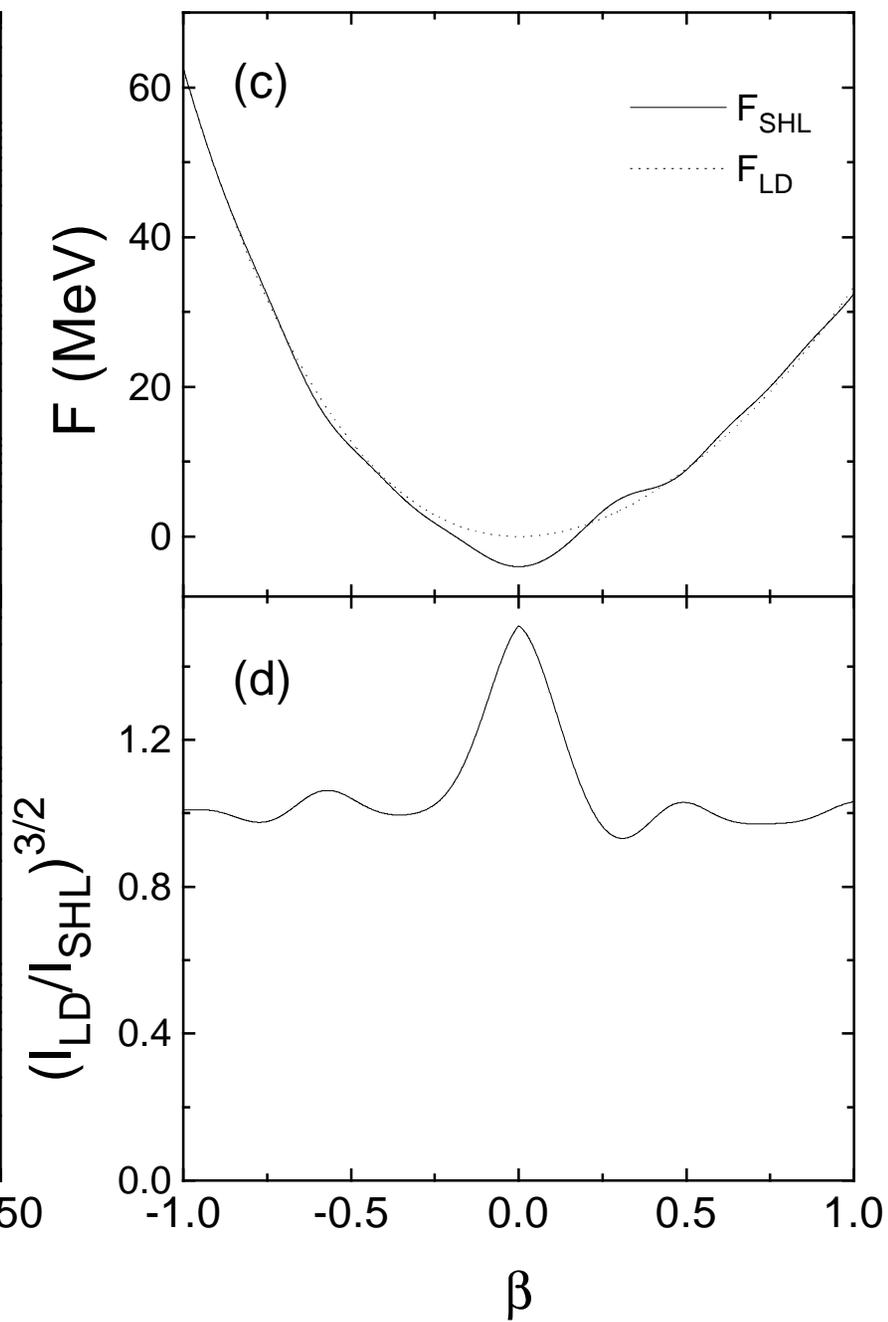
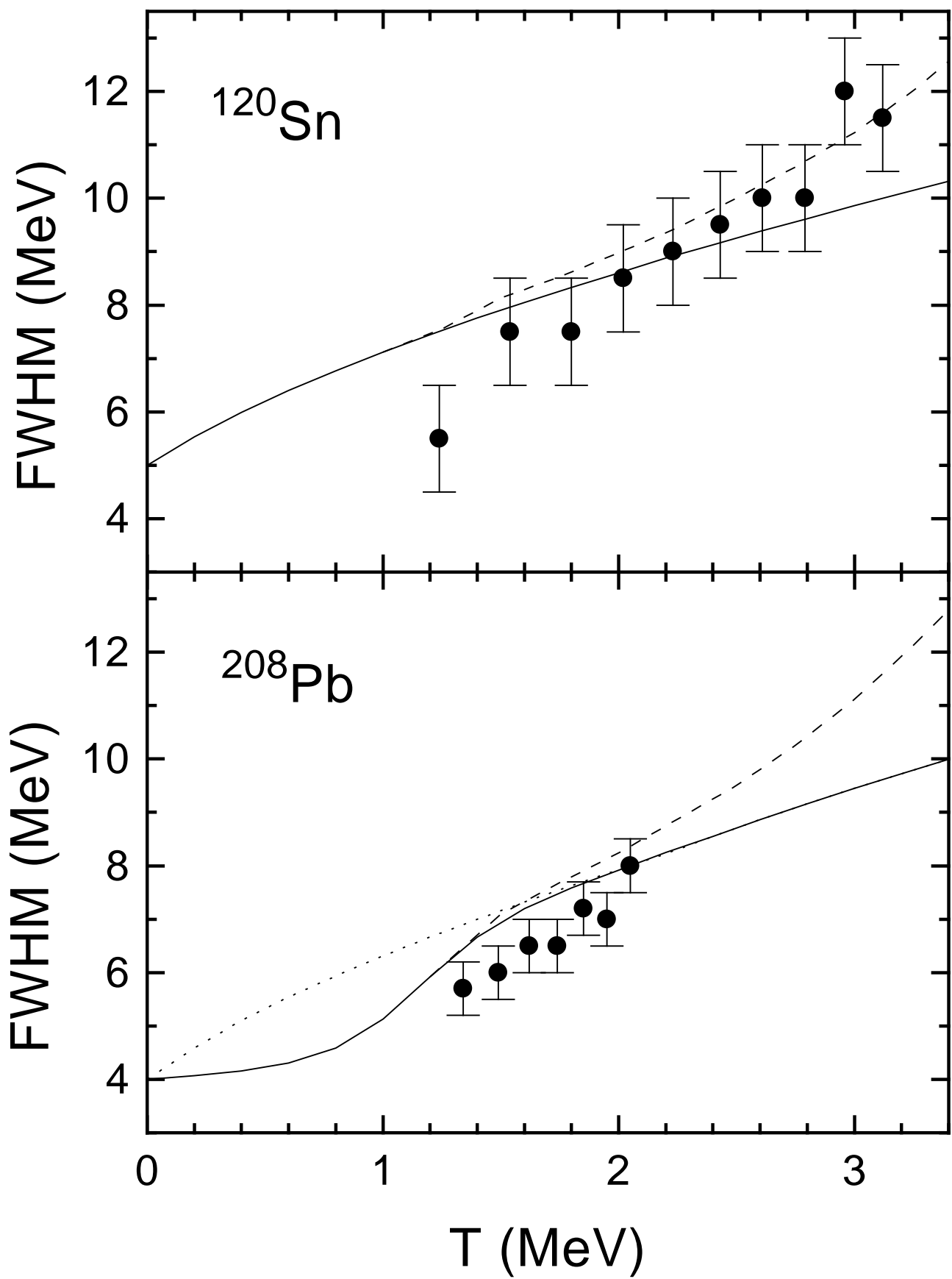
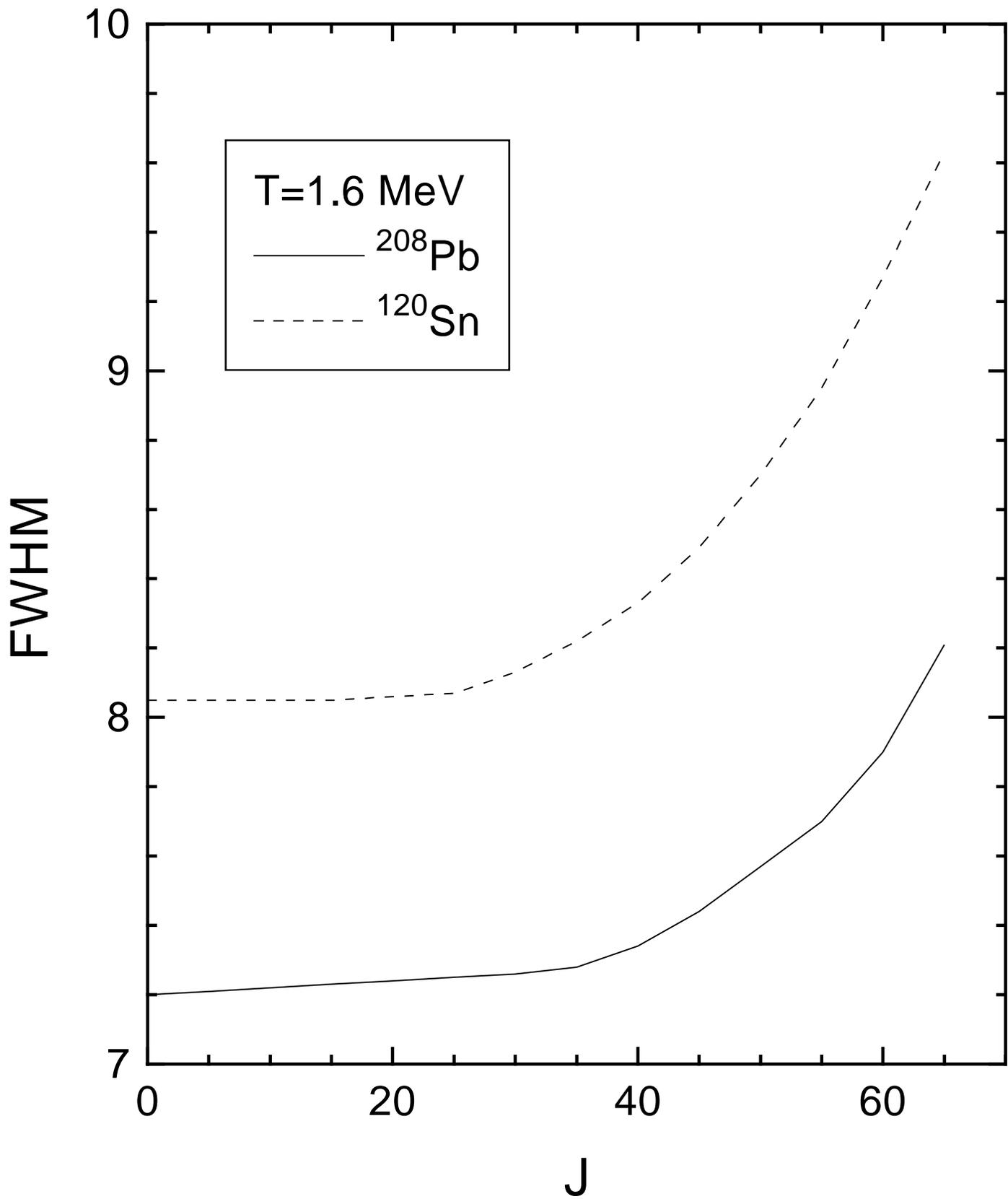
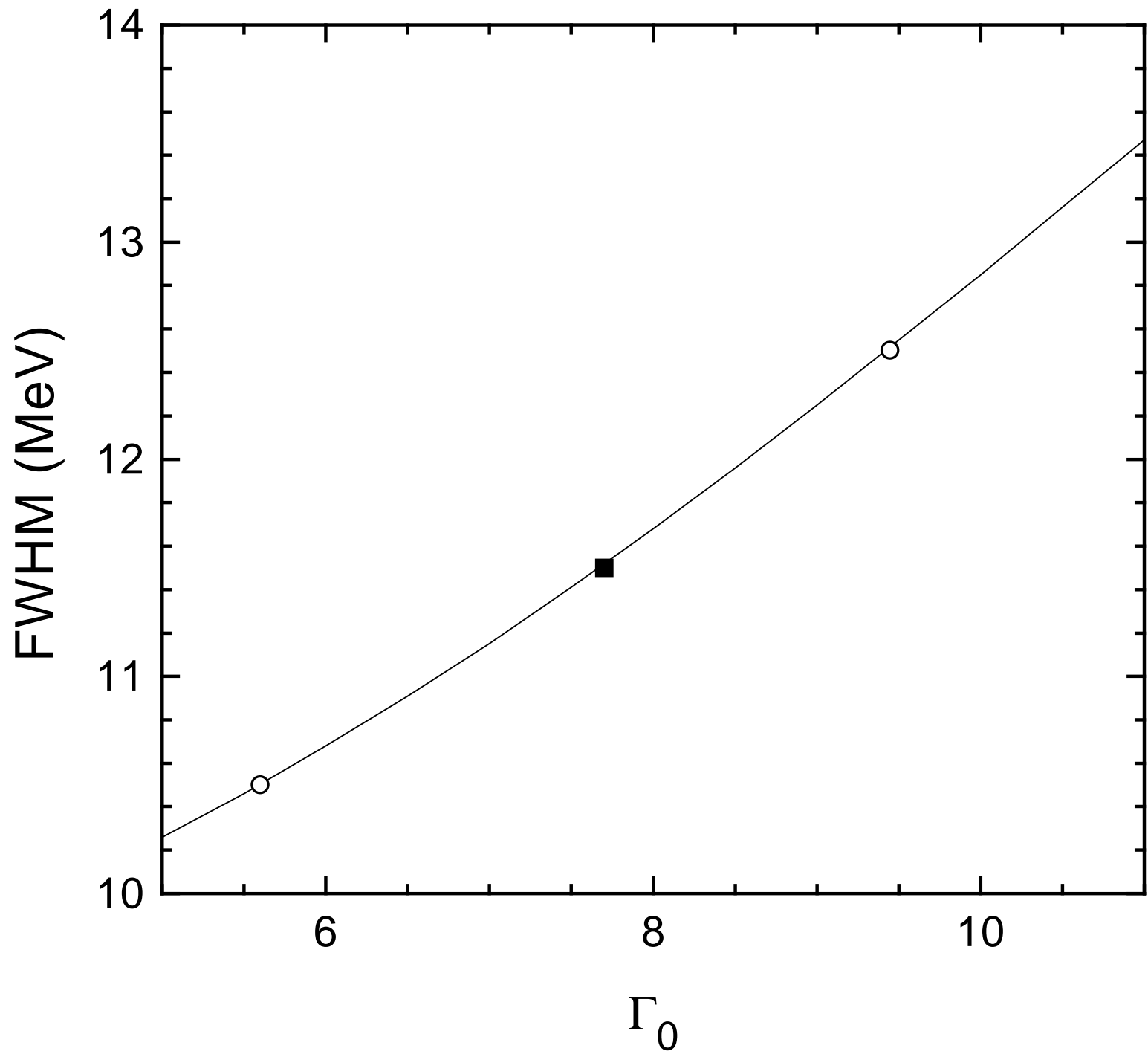
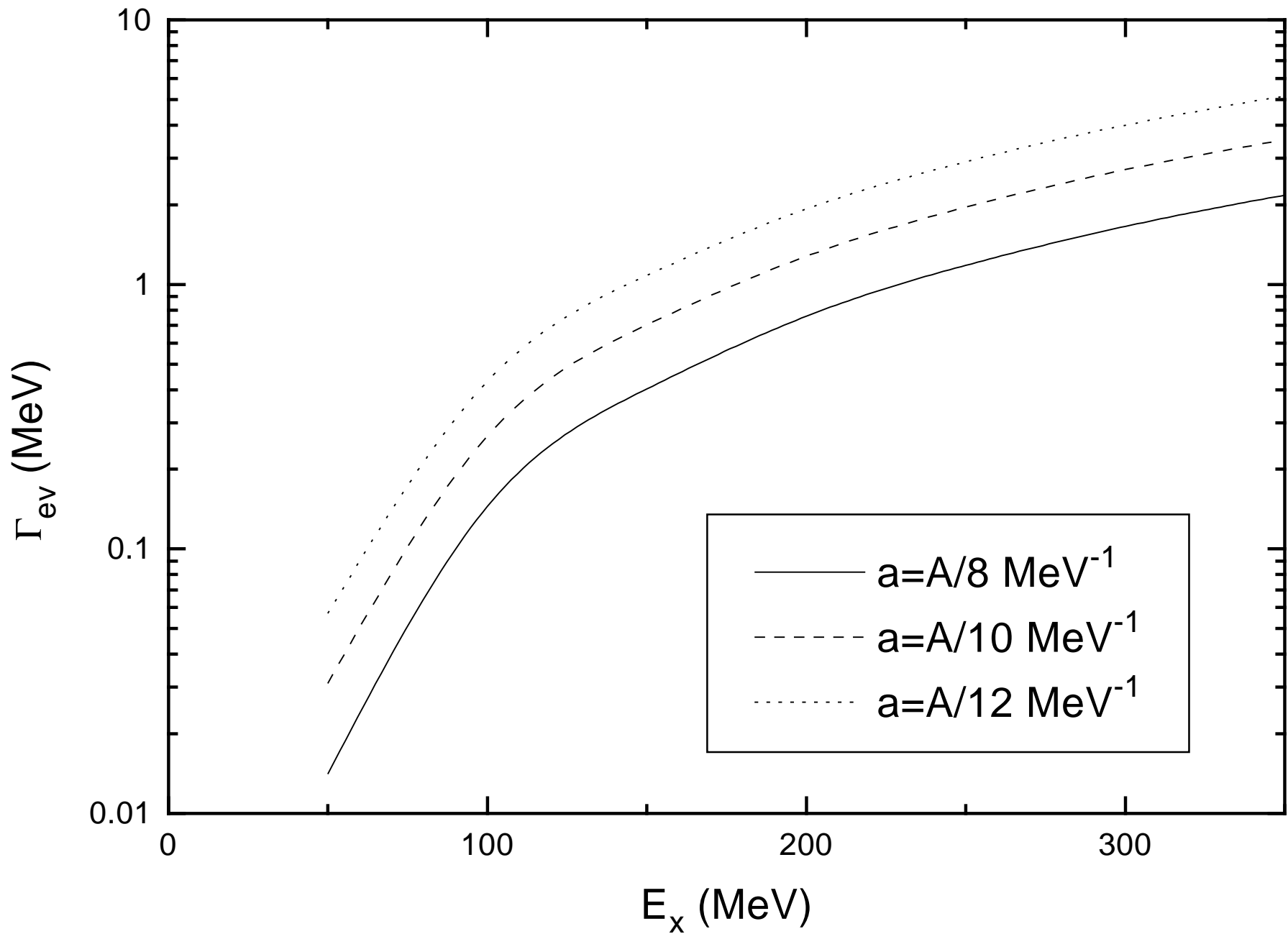
\end{document}